\documentclass[apjl]{emulateapj}
\usepackage{rotate}
\usepackage{amsmath}
\usepackage{natbib}
\bibpunct{(}{)}{;}{a}{}{,}

\usepackage{graphicx,psfig,epsf}  
\usepackage{times}
\usepackage{subfigure}
\usepackage{textcomp}
\usepackage{multirow}
\usepackage{amsmath}

\def\apjl{ApJ}
\def\aj{AJ}                   
\def\araa{ARA\&A}             
\def\apj{ApJ}                 
\def\apjl{ApJ}                
\def\apjs{ApJS}               
\def\ao{Appl.~Opt.}           
\def\aap{A\&A}                
\def\mnras{MNRAS}             
\def\pasj{PASJ}               
\def\nat{Nature}              

\def\xmm {\emph{XMM-Newton}}
\def\nus {\emph{NuSTAR}}

\received{}
\revised{}
\accepted{}

\begin{document}

\shorttitle{The search for new ultraluminous pulsars}
\shortauthors{Pintore et al.}

\title{Pulsator-like spectra from Ultraluminous X-ray sources and the search for more ultraluminous pulsars}

\author{F. Pintore\altaffilmark{1}, L. Zampieri\altaffilmark{2}, L. Stella\altaffilmark{3}, A. Wolter\altaffilmark{4}, S. Mereghetti\altaffilmark{1},   G. L. Israel\altaffilmark{3} }

\altaffiltext{1}{INAF-IASF Milano, via E. Bassini 15, I-20133 Milano, Italy}
\altaffiltext{2}{NAF-Osservatorio Astronomico di Padova, Vicolo dell'Osservatorio 5, I-35122 Padova, Italy}
\altaffiltext{3}{INAF - Osservatorio astronomico di Roma, Via Frascati 44, I-00078, Monteporzio Catone, Italy}
\altaffiltext{4}{INAF, Osservatorio Astronomico di Brera, via Brera 28, 20121 Milano, Italy}

\begin{abstract}

Ultraluminous X-ray sources (ULXs) are a population of extragalactic objects whose luminosity exceeds the Eddington limit for a 10 M$_\odot$ black hole (BH). Their properties have been widely interpreted in terms of accreting stellar-mass or intermediate-mass BHs. However at least three neutron stars (NSs) have been recently identified in ULXs through the discovery of periodic pulsations. Motivated by these findings we studied the spectral properties of a sample of bright ULXs using a simple continuum model which was extensively used to fit the X-ray spectra of accreting magnetic NSs in the Galaxy. We found that such a model, consisting of a power-law with a high-energy exponential cut-off, fits very well most of the ULX spectra analyzed here, at a level comparable to that of models involving an accreting BH. On these grounds alone we suggest that other non-pulsating ULXs may host NSs. We found also that above 2 keV the spectrum of known pulsating ULXs is harder than that of the majority of the other ULXs of the sample, with only IC 342 X-1 and Ho IX X-1 displaying spectra of comparable hardness. We thus suggest that these two ULXs may host an accreting NS and encourage searches for periodic pulsations in the flux.

\end{abstract}


\section{Introduction}

Ultraluminous X-ray sources (ULXs) are a class of extragalactic, point-like sources with  luminosity from $\sim10^{39}$ erg s$^{-1}$ up to $10^{42}$ erg s$^{-1}$, in excess of the Eddington limit for a 10$M_{\odot}$ black hole (BH)  (e.g. \citealt{fabbiano06}). Such a  high X-ray luminosity suggests that they host either  intermediate mass BHs  accreting at sub-Eddington rate \citep{colbert99,miller04}, or super-Eddington accreting BHs of stellar mass or of stellar origin (e.g. \citealt{gladstone09,fengsoria11}). 
In particular,  the most luminous ULXs ($>5\times10^{40}$ erg s$^{-1}$) are usually considered as the best intermediate mass BH candidates \citep[e.g.][]{sutton12}, while less luminous sources could be super-Eddington accreting BHs with masses in the range 5--80 M$_\odot$ \citep[e.g.][]{zampieri09}.
The recent discovery of three pulsating neutron stars (NSs) in the ULX M82 X-2 \citep{bachetti14}, NGC 7793 P13 and NGC 5907 X-1 (\citealt{israel16b}, \citealt{israel16a}) proved  that ULXs can be powered by objects different from  BHs.

The ULXs spectra are generally described in terms of  a cold (kT $\sim0.1-0.6$ keV) soft thermal component coupled to a high-energy component with a cut-off below 10 keV \citep[e.g.][]{sutton13,bachetti13,pintore14,middleton15}. The soft component is possibly associated to the photosphere of strong outflows ejected from the accretion disc at super-Eddington rates  \citep[e.g.][]{poutanen07,ohsuga11}, while the high-energy component might come from the bare inner accretion disc or from an optically-thick corona close to the BH \citep[e.g.][]{middleton15}. However, the quality of ULX spectra is generally low and the interpretation of their spectral properties is still matter of debate.

By exploiting the recent availability of broadband X-ray data obtained with the \xmm\ and \nus\ satellites, we carried out a comprehensive spectral analysis  of a large sample of  the brightest  ULXs. Several previous investigations focussed on the phenomenological characterization of the spectra of ULXs \citep[e.g][]{goncalves06,stobbart06,gladstone09,pintore12,sutton13,pintore14}. 
Motivated by the discovery of X-ray pulsations from M82 X-2 and subsequently from NGC 7793 P13 and NGC 5907 X-1, we adopted a simple spectral model that was extensively used to fit the X-ray continuum of accreting magnetic NSs, especially X-ray pulsators in Galactic high mass X-ray binaries \citep[for more details see][]{white83, white95, coburn01}.
The model consists of a power-law with a high-energy exponential cut-off plus a blackbody; we find that in most cases the ULX spectra are well fit by such a model, indicating that different interpretations of the ULX spectra are possible.

We describe the data reduction in section~\ref{data_reduction}, we present the results of our spectral analysis in section~\ref{spectral} and we discuss them in section~\ref{discussion}.

\section{Observations and data analysis}
\label{data_reduction}

\begin{table*}
  \begin{center}
   \caption{Log of the observations of the ULXs analyzed in this work and ordered with increasing RA.}
\footnotesize
   \label{log}
   \begin{tabular}{l l l r l l c l}
\hline
Source. &Obs.ID & Date & Exp. & Obs.ID &  Date &Exp & Epoch\\
&  \xmm\  & & (ks) & \nus\  & &(ks) \\
\hline
\multirow{3}{*}{NGC 5907 X-1} & 0724810401 & 2013-11-12 & 33.5 & 30002039005 & 2013-11-12 & 113  & 1\\
  & 0729561301 & 2014-07-09 & 42.0 & 80001042002 & 2014-07-09 & 57.1  & 2\\
 & - & - & - & 80001042004 & 2014-07-12 & 56.3 & 2 \\
\hline
\multirow{2}{*}{NGC 7793 P13} & 0693760401 & 2013-11-25 & 46.0 & - & - & -  & 1\\
 & 0748390901 & 2014-12-10 & 47.0 & - & - & - & 2 \\
\hline
\hline
\multirow{2}{*}{NGC 55 ULX-1} & 0028740101 & 2001-11-15 & 28.3 & - & - & - & 1\\
 & 0655050101 & 2010-05-24 & 124.0 & - & - & - & 2\\
\hline
\multirow{2}{*}{NGC 1313 X-1} & 0150280301 & 2003-12-21 & 10.3 & - & - & -  & 1\\
 & 0693851201 & 2012-12-22 & 125.2 & 30002035004 & 2012-12-16 &127.0 & 2 \\
\hline
\multirow{2}{*}{NGC 1313 X-2} & 0405090101 & 2012-12-16 & 123.3 & - & - & - & 1 \\
 & 0693851201 & 2012-12-22 & 125.2 & 30002035004 & 2012-12-16 &127.0 & 2 \\
\hline
\multirow{2}{*}{IC 342 X-1} & 0693850601 & 2012-08-11 & 59.9 & 30002032003 & 2012-08-10 & 98.6 & 1 \\
 & 0693851301 & 2012-08-17 & 60.0 & 30002032005 & 2012-08-16 & 127.4 & 2 \\
\hline
\multirow{4}{*}{Ho II X-1} & 0200470101 & 2004-04-15 & 93.5 & - & - & - & 1 \\
& 0724810101 & 2013-09-09 & 12.3 & 30001031002 & 2013-09-09 & 31.4 & 2 \\
 & - & - & - & 30001031003  & 2013-09-09 & 79.4 & 2 \\
 & 0724810301 & 2013-09-17 & 12.0 & 30001031005  & 2013-09-17& 111.1 & 2 \\
\hline
\multirow{5}{*}{Ho IX X-1} & 0200980101 & 2004-09-26 & 117  & - & - & - & 1\\
 & - & - & - & 30002033005  & 2012-11-11 & 40.7 & 2\\
 & 0693851701 & 2012-11-12 & 9.92 & 30002033006   & 2012-11-11 &35.2 & 2 \\
 & 0693851801 & 2012-11-14 & 13.8 & 30002033008  & 2012-11-14 &14.5 & 2 \\
 & 0693851101 & 2012-11-16 & 13.3  & 30002034010  & 2012-11-15 &49.0 & 2 \\
\hline
\multirow{2}{*}{NGC 5204 X-1} & 0405690201 & 2006-11-19 & 43.5 & - & - & - & 1\\
& 0693851401 & 2013-04-21 & 16.9 & 30002037002 & 2013-04-19&96.0 & 2 \\
\hline
\multirow{1}{*}{NGC 5408 X-1} & 0653380501 & 2011-01-28 & 124.1 & - & - &-  & 1\\
\hline
\multirow{1}{*}{NGC 5643 ULX-1} & 0744050101 & 2014-08-27 & 114.0 & - & - & - & 1\\
\hline
\multirow{1}{*}{NGC 6946 X-1} & 0691570101 & 2012-10-21 & 114.3 & - & - & - & 1\\
\hline
\end{tabular}
\end{center}
\end{table*}
{ We analyzed a sample of 12 ULXs, including two ULX pulsars, located at distances lower than 15 Mpc, which shows a broad luminosity range spanning from $\sim10^{39}$ erg s$^{-1}$ to $\sim10^{41}$ erg s$^{-1}$. The sources have available \xmm\ and, in some cases, simultaneous (or nearly simultaneous) \nus\ observations of good quality which allow us to put robust constraints on the high energy curvature of their X-ray spectrum (i.e. with at least 10000 counts in the {\it XMM-Newton} EPIC instruments, as shown in \citealt{gladstone09}). Our sample is similar to those analyzed in several previous studies \citep[e.g.][]{stobbart06,gladstone09,pintore14,middleton15} and offers an overall picture of the  spectral properties displayed by ULXs.}
In order to extract individual spectra with better statistics, observations of the same source obtained within time intervals of a few days were merged if no significant spectral variations were present. 
Data taken at very different times and/or showing different spectral states were instead considered as individual observations. { Additionally to this selection, we also considered, for each source, the observations with the largest source spectral variability.}
We list in Table~\ref{log} the observations that are used in this work.

The data of the \xmm\  EPIC instrument \citep{struder01short,turner01short} were reduced using SAS v.14.0.0. We extracted spectra from events with {\sc pattern} $\leq 4$ for EPIC-pn (single- and double-pixel events) and {\sc pattern} $ \leq 12$ for EPIC-MOS (single- and multiple-pixel events). We set `{\sc flag}=0' in order to exclude bad pixels and events coming from the CCD edges and removed time intervals with  high background.  The spectra were extracted from circular regions with radius of 30$''$ and 65$''$   for source and background, respectively.

The \nus\ data were reduced with the standard pipeline, based on the \nus\  {\it  Data Analysis Software} v1.3.0 ({\sc NUSTARDAS}) in the HEASOFT {\sc ftools} v6.16. We obtained cleaned event files by applying the standard corrections. Spectra were extracted from circular   regions of radius 40$''$ and 60$''$ for source and background, respectively.

{ All {\it XMM-Newton} and {\it NuSTAR} spectra were rebinned so as to have at least 25 and 100 counts per bin, respectively, so that minimum $\chi^2$ fitting techniques could be used}; model fitting was carried out using XSPEC v.12.8.2 \citep{arnaud96}. The spectra of the pn and the two MOS cameras, and (when available) of the \nus\ detectors were fit together, for each epoch. A multiplicative factor was included to account for possible inter-calibration uncertainties, which were in most cases smaller than 10\%. We considered the 0.3--10 keV and 3--70 keV energy range for EPIC and \nus data, respectively.

\section{Results}
\label{spectral}
\subsection{Broadband spectral analysis}

We fit the spectra with an absorbed power-law   with an exponential  cut-off characterized by two parameters, the cut-off energy $E_c$ and the folding energy $E_f$: $F(E)=k\cdot E^{-\Gamma} e^{-(E-E_c)/E_f  }$ ({\sc highecut$\times$powerlaw} in {\sc xspec}), with the exponential part acting for $E>E_c$.
In most cases,  an additional  soft component, that we  model with a blackbody for simplicity, is needed. 
This model was extensively applied in the past to fit the X-ray continuum of accreting magnetic NSs, notably X-ray pulsators in high mass X-ray binaries. In those systems the hard component is believed to originate in a magnetic accretion-column close to the NS surface, whereas the soft blackbody-like component is usually ascribed to emission from the inner edge of the disk at the magnetospheric boundary  \citep[e.g.][]{white95, coburn01}. 

NGC 5907 X-1 was observed six times with \xmm\ (e.g. \citealt{israel16a}, \citealt{fuerst16}) and good quality spectra are available for the observations of February 2003, February 2012, November 2013 and July  2014.    Simultaneous and quasi-simultaneous \nus\  data are available for the 2013 and 2014  observations.  The fit of the 2003, 2012 and 2014 spectra give results consistent within the errors with absence of spectral variability, as also found by \citet{fuerst16}. We therefore report here only the results from the 2014 observations which are of higher quality and can be complemented with \nus\ data. They can be fit by a {\sc highecut$\times$powerlaw} alone (no soft component) with photon index $\Gamma$$\sim$1.5,   E$_c$ = 5.5 keV and E$_f$ = 8.3 keV.    
The 2013 spectrum, obtained when the source was about a factor of 2-3 fainter than in 2014,  is slightly softer ($\Gamma\sim$1.9), but it can still be fit by a   cut-off power-law without the addition of a soft  component.

For  NGC 7793 P13, we use two observations taken  in 2013 and 2014 (no \nus\  data are available).
Both observations  give similar results,   requiring  a thermal component with  temperature kT$\sim$0.2 keV in addition to a cut-off power-law  with $\Gamma$$\sim$1.1.    Another \xmm\  observation, obtained  in May 2012 when the source was much fainter,  provides a less constraining spectrum, but consistent in spectral shape with those of 2013 and 2014.

\begin{figure}
\center
\includegraphics[height=14.6cm]{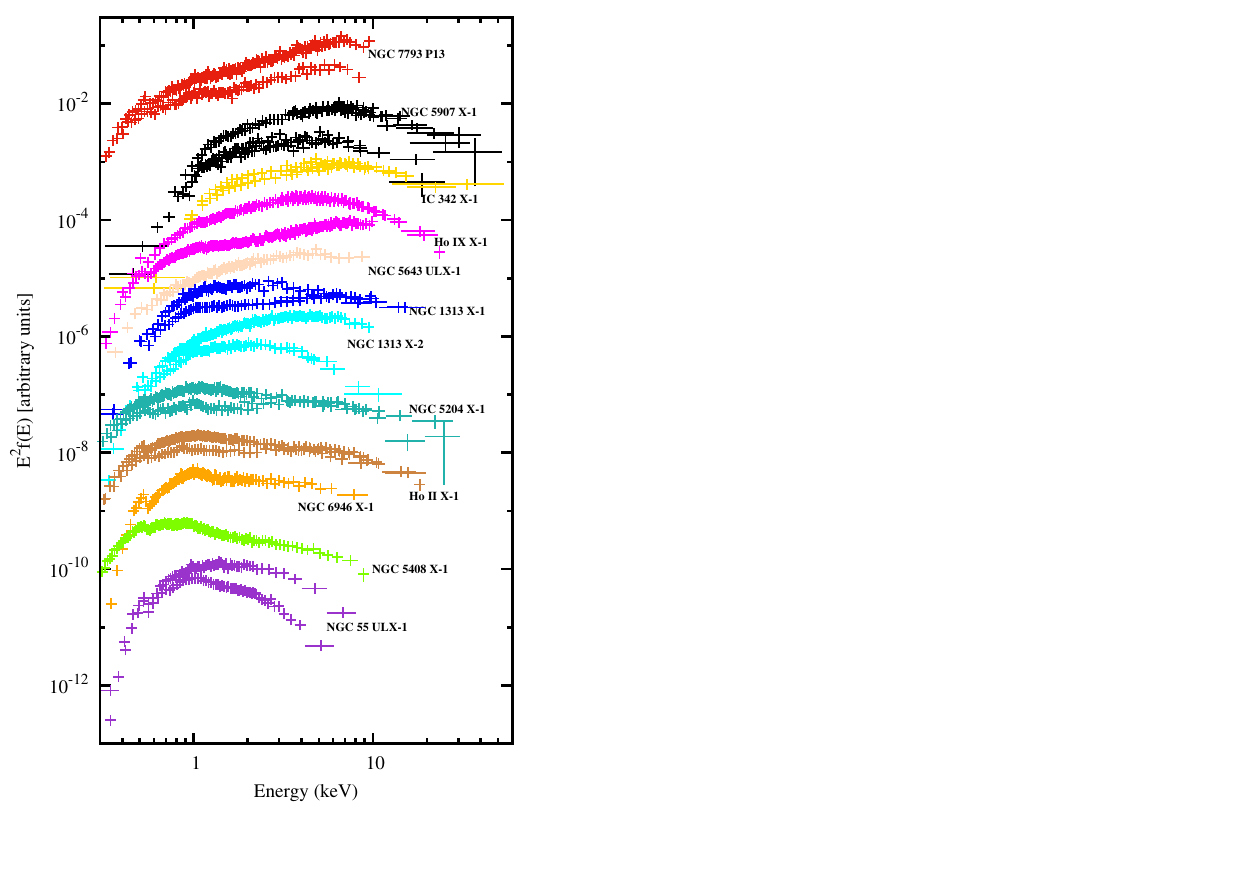}
\vspace{-1.5cm}
\caption{Unfolded (E$^2$F(E)) spectra of all the ULXs of our sample. For each source, we show the two most different spectral states obtained by fitting them with the model for accreting NSs described in the text. For display purposes, the spectra have been rebinned only and the flux is arbitrarily shifted on the {\it y} axis.}
\label{fig_spectra}
\end{figure}

For all other ULXs in our sample, several \xmm\ and \nus\  observations are available. We analyze all the available good quality spectra, finding in most cases evidence for spectral variability, as already reported in the literature \citep[e.g.][]{sutton13,pintore14,middleton15}.
To illustrate the range of possible states in which the sources can be found, we plot in Fig.~\ref{fig_spectra} the  two representative spectra  of each ULX,  choosing for each source the observations with the hardest and softest spectrum (see section~\ref{data_reduction}).
These are the observations reported in Table~\ref{log}, with corresponding  best fit  parameters listed in Table~\ref{spectr}.  

\begin{table*}
  \begin{center}
   \caption{Spectral results obtained by adopting the continuum model {\sc tbabs$\times$(bbody + highecut$\times$powerlaw)}.} 
    \scalebox{0.74}{\begin{minipage}{25.5cm}
\footnotesize
   \label{spectr}
   \begin{tabular}{l c l l l l l l l l l l c r}
\hline
Src. & Epoch &\multicolumn{1}{c}{{\sc TBabs}} & \multicolumn{2}{c}{{\sc bbody}} & \multicolumn{2}{c}{{\sc highecut}}  & \multicolumn{2}{c}{{\sc powerlaw}} & \multicolumn{3}{c}{\sc gabs} & \multicolumn{1}{c}{} & \multicolumn{1}{c}{} \\
 & &\multicolumn{1}{c}{N$_H$} & \multicolumn{1}{c}{kT$_{in}$} & \multicolumn{1}{c}{Norm} & \multicolumn{1}{c}{E$_c$} & \multicolumn{1}{c}{ E$_f$}   & \multicolumn{1}{c}{$\Gamma$}  & \multicolumn{1}{c}{Norm} & Energy & $\sigma$ & Strenght & \multicolumn{1}{c}{L$_\text{X}$$^a$} & \multicolumn{1}{c}{$\chi^2/dof$} \\
& &\multicolumn{1}{c}{10$^{22}$ cm$^{-2}$} & \multicolumn{1}{c}{keV} & \multicolumn{1}{c}{$10^{-6}$} & \multicolumn{1}{c}{keV} & \multicolumn{1}{c}{keV}   & \multicolumn{1}{c}{}  & \multicolumn{1}{c}{10$^{-3}$} & keV & keV &  & \multicolumn{1}{c}{$10^{40}$ erg s$^{-1}$} & \\
\hline
 \hline
 \multirow{2}{*}{NGC 5907 X-1} & 1 &  $0.76^{+0.07}_{-0.06}$ & - & - & $6.0^{+0.9}_{-1.5}$ & $8.0^{+2.5}_{-1.8}$ & $1.85^{+0.08}_{-0.09}$ & $0.19^{+0.02}_{-0.02}$ & - & - & - & 4.2 & 149.44/173\\
 & 2 &  $0.77^{+0.04}_{-0.04}$  & -  & -  & $5.5^{+0.5}_{-0.5}$ & $8.3^{+0.8}_{-0.7}$ &   $1.48^{+0.04}_{-0.05}$ &   $0.34^{+0.02}_{-0.02}$ & - & - & - & 10.9 & 442.90/442 \\
\multirow{2}{*}{NGC 7793 P13}  & 1 & $0.07^{+0.02}_{-0.02}$ & $0.22^{+0.02}_{-0.02}$ & $1.5^{+0.3}_{-0.2}$ & $5.3^{+0.4}_{-1.2}$ & $4.3^{+1.7}_{-0.8}$ & $1.06^{+0.08}_{-0.09}$ &  $0.093^{+0.009}_{-0.01}$ & - & - & - & 0.21 &  357.49/325\\
  & 2 & $0.10^{+0.02}_{-0.02}$ & $0.19^{+0.04}_{-0.03}$ & $1.1^{+0.5}_{-0.4}$ & $6.6^{+0.3}_{-1.2}$ & $4.5^{+3.4}_{-1.1}$ & $1.11^{+0.04}_{-0.07}$ & $0.24^{+0.01}_{-0.01}$ & - & - & - & 0.53 & 629.78/563\\
\hline
\hline
\multirow{2}{*}{NGC 55 ULX-1}  & 1 & $0.291^{+0.051}_{-0.044}$ & $0.173^{+0.022}_{-0.02}$ & $29.3^{+5.8}_{-5.2}$ & $1.424^{+0.12}_{-0.09}$ & $1.48^{+0.44}_{-0.32}$ & $0.95^{+0.51}_{-0.64}$ & $8.1^{+3}_{-2.8}$ & - & - & - & 0.26 & 700.77/688 \\
  & 2 & $0.41^{+0.06}_{-0.05}$ & $0.15^{+0.01}_{-0.01}$ & $4.0^{+0.6}_{-0.7}$ & $2.00^{+0.1}_{-0.09}$ & $1.5^{+0.4}_{-0.3}$ & $2.3^{+0.4}_{-0.5}$ & $5.9^{+1.9}_{-1.8}$ & $0.75^{+0.01}_{-0.01}$ & $0.07^{+0.02}_{-0.01}$ & $0.038^{+0.01}_{-0.009}$ & 0.23 & 270.40/243  \\

\\
\multirow{2}{*}{NGC 1313 X-1}   & 1 & $0.3^{+0.1}_{-0.1}$ & $0.24^{+0.06}_{-0.08}$ & $19^{+12}_{-14}$ & $2.9^{+0.7}_{-2.9}$ & $4.6^{+7.2}_{-2.5}$ & $1.7^{+0.5}_{-1.2}$ & $1.1^{+0.8}_{-0.8}$ & - & - & - & 1.67 & 250.54/229\\
  & 2 & $0.283^{+0.01}_{-0.008}$ & $0.205^{+0.008}_{-0.008}$ & $9.4^{+0.6}_{-0.6}$ & $6.1^{+0.5}_{-0.9}$ & $10.4^{+1.4}_{-1.1}$ & $1.74^{+0.03}_{-0.06}$ & $5.8^{+0.3}_{-0.4}$ & - & - & - & 1.07 & 1703.40/1718  \\
\\
\multirow{2}{*}{NGC 1313 X-2}  & 1 & $0.32^{+0.01}_{-0.01}$ & - & - & $3.4^{+0.4}_{-0.3}$ & $7.7^{+1}_{-0.9}$ & $1.60^{+0.04}_{-0.04}$ & $0.68^{+0.02}_{-0.02}$  & - & - & - & 1.05 & 820.64/860 \\
   & 2 & $0.19^{+0.04}_{-0.04}$ & $0.23^{+0.03}_{-0.04}$ & $1.7^{+1.1}_{-1.1}$ & $1.02^{+0.06}_{-0.06}$ & $1.8^{+0.4}_{-0.3}$ & $0.9^{+0.4}_{-0.4}$ & $0.28^{+0.04}_{-0.06}$ & - & - & - & 0.28 & 1020.57/987  \\

\\

\multirow{2}{*}{IC 342 X-1}  & 1 & $1.045^{+0.087}_{-0.075}$ & $0.219^{+0.028}_{-0.024}$ & $8.7^{+4.1}_{-2.5}$ & $6.66^{+0.45}_{-0.56}$ & $10.6^{+1.4}_{-1.2}$ & $1.62^{+0.056}_{-0.065}$ & $0.51^{+0.04}_{-0.05}$ & - & - & - & 0.43 & 937.17/981 \\
  & 2 & $0.99^{+0.026}_{-0.026}$ & - & - & $6.38^{+0.5}_{-0.67}$ & $15^{+2.4}_{-1.9}$ & $1.915^{+0.03}_{-0.031}$ & $0.89^{+0.03}_{-0.03}$ & - & - & - & 0.51 & 1061.76/1063 \\

\\
\multicolumn{1}{l}{Ho II X-1}  & 1 & $0.12^{+0.01}_{-0.01}$ & $0.211^{+0.006}_{-0.007}$ & $2.0^{+0.2}_{-0.3}$ & $2.0^{+0.2}_{-0.3}$ & $8.7^{+2.4}_{-1.6}$ & $2.1^{+0.1}_{-0.1}$ & $1.6^{+0.1}_{-0.1}$ & - & - & - & 1.2 & 1317.83/1317\\
  & 2 &  $0.10^{+0.02}_{-0.02}$  & $0.17^{+0.01}_{-0.01}$ & $13.1^{+2.2}_{-2.0}$  & $5.3^{+0.3}_{-0.5}$ & $8.6^{+0.7}_{-0.6}$ & $1.93^{+0.06}_{-0.08}$ & $1.00^{+0.07}_{-0.08}$ & - & - & - & 0.88 & 800.29/785\\

\\
\multirow{2}{*}{Ho IX X-1}  & 1 & $0.160^{+0.008}_{-0.008}$ & $0.209^{+0.008}_{-0.008}$ & $8.7^{+0.5}_{-0.5}$ & $6.0^{+0.4}_{-0.5}$ & $10.6^{+2.6}_{-2.1}$ & $1.38^{+0.03}_{-0.03}$ & $0.70^{+0.02}_{-0.02}$ & - & - & - & 1.4 & 2191.03/2043\\
 & 2 & $0.239^{+0.008}_{-0.008}$ &$0.21^{+0.06}_{-0.05}$ & $2.1^{+0.9}_{-0.9}$ & $3.6^{+0.2}_{-0.2}$ & $5.7^{+0.2}_{-0.2}$ & $1.43^{+0.04}_{-0.04}$ & $2.8^{+0.1}_{-0.1}$ & - & - & - & 3.8 &1546.80/1433\\

\\
\multirow{2}{*}{NGC 5204 X-1}  & 1 & $0.08^{+0.02}_{-0.03}$ & $0.240^{+0.009}_{-0.02}$ & $6.9^{+1.3}_{-1.2}$ & $2.8^{+0.5}_{-2.8}$ & $12.4^{+12}_{-5.4}$ & $2.0^{+0.1}_{-0.3}$ & $0.45^{+0.03}_{-0.08}$ & - & - & - & 0.81 & 837.88/821 \\
  & 2 & $0.05^{+0.03}_{-0.03}$ & $0.18^{+0.02}_{-0.02}$ & $3.6^{+1.0}_{-0.8}$ & $4.0^{+0.7}_{-1.1}$ & $7.4^{+1.5}_{-1.7}$ & $1.7^{+0.1}_{-0.3}$ & $0.23^{+0.03}_{-0.05}$ & - & - & - & 0.50 & 414.06/414  \\

\\
\multirow{2}{*}{NGC 5408 X-1}  & \multirow{2}{*}{1} & \multirow{2}{*}{$0.09^{+0.01}_{-0.01}$} & \multirow{2}{*}{$0.162^{+0.005}_{-0.005}$}  & \multirow{2}{*}{$14.8^{+1.5}_{-1.6}$} & \multirow{2}{*}{$2.0^{+0.3}_{-0.2}$} & \multirow{2}{*}{$5.7^{+2.3}_{-1.4}$} & \multirow{2}{*}{$2.0^{+0.2}_{-0.2}$} & \multirow{2}{*}{$0.32^{+0.04}_{-0.05}$} & $0.75^{+0.01}_{-0.02}$ & $0.07^{+0.03}_{-0.03}$ & $0.03^{+0.02}_{-0.01}$ & \multirow{2}{*}{0.87} & \multirow{2}{*}{1035.47/990} \\
 &  &  &  & &  & & & & $1.17^{+0.03}_{-0.03}$ & $0.08^{+0.04}_{-0.03}$ & $0.023^{+0.02}_{-0.009}$ &  &  \\

\\
\multirow{1}{*}{NGC 5643 ULX-1}  & 1 & $0.20^{+0.01}_{-0.01}$ & - & - & $3.9^{+0.3}_{-0.3}$ & $6.5^{+1.1}_{-0.9}$ & $1.46^{+0.04}_{-0.04}$ & $0.160^{+0.005}_{-0.005}$ & - & - & - & 2.98 & 865.07/934 \\

\\
\multirow{1}{*}{NGC 6946 X-1}  & 1 & $0.40^{+0.07}_{-0.06}$ & $0.13^{+0.01}_{-0.02}$ & $25^{+18}_{-8}$ & $1.8^{+0.3}_{-0.4}$ & $7.0^{+4.7}_{-2.3}$ & $1.9^{+0.2}_{-0.3}$ & $0.25^{+0.05}_{-0.04}$ & $0.69^{+0.03}_{-0.04}$ & $0.13^{+0.04}_{-0.03}$ & $0.15^{+0.13}_{-0.05}$ & 0.9 & 613.68/637 \\

 \hline
\end{tabular}
\flushleft $^a$ Unabsorbed luminosity in the 0.3--20 keV energy band; the adopted distances are: 17.14 Mpc, 3.58 Mpc, 2.73 Mpc, 3.77 Mpc, 3.27 Mpc, 5.32 Mpc, 4.25 Mpc, 1.99 Mpc and 4.76 Mpc \citep{tully13} for \\NGC 5907, NGC 7793, IC 342, Ho IX, Ho II, NGC 5408, NGC 1313, NGC 55 and NGC 5204, respectively; 5.5 Mpc \citep{smith07} for NGC 6946 and 13.9 Mpc \citep{sanders03} for NGC 5643.
\end{minipage}}
\end{center}
\end{table*}

Generally, the ``pulsator-like'' model gives good fits, except for a few cases in which the soft thermal component is not strictly required (i.e.  IC 342 X-1  in August 2012,  and NGC 1313 X-2  in December 2012). 
In all other cases, blackbody components with temperatures in the range $\sim$0.15--0.3 keV are required.
A variety of power-law spectral slopes and cut-off parameters is found (Table~\ref{spectr}). 

The relatively large $\chi^2$ values obtained in the fit of Ho IX X-1  are caused by some  excess emission   at  high energy,  possibly indicating the presence of a further hard tail component,  as discussed in \citet{walton14b}. { We note that hard tails are observed also in the spectra of some Galactic accreting NSs \citep[e.g.][]{mcclintock06}. } 
  
{ Finally, the spectra of  NGC 55 ULX1, NGC 5408 X-1, and NGC 6946 show some residuals  around 1 keV, as already  reported in literature \citep[e.g.][]{middleton14}. These features have been associated lately to a blending of unresolved emission and absorption lines produced by the photosphere of a possible outflowing wind in a scenario of super-Eddington accretion on a stellar mass compact object \citep{pinto16}.}
For the three sources, we find that good fits are obtained   with the addition  of one or two broad Gaussian absorption lines centered at $\sim$0.7 keV and $\sim$1.2 keV and with widths  $\sigma\sim0.07-0.13$ keV.

\subsection{Characterization of the hard component}

\begin{figure*}
\center
\includegraphics[width=15.0cm,height=11cm]{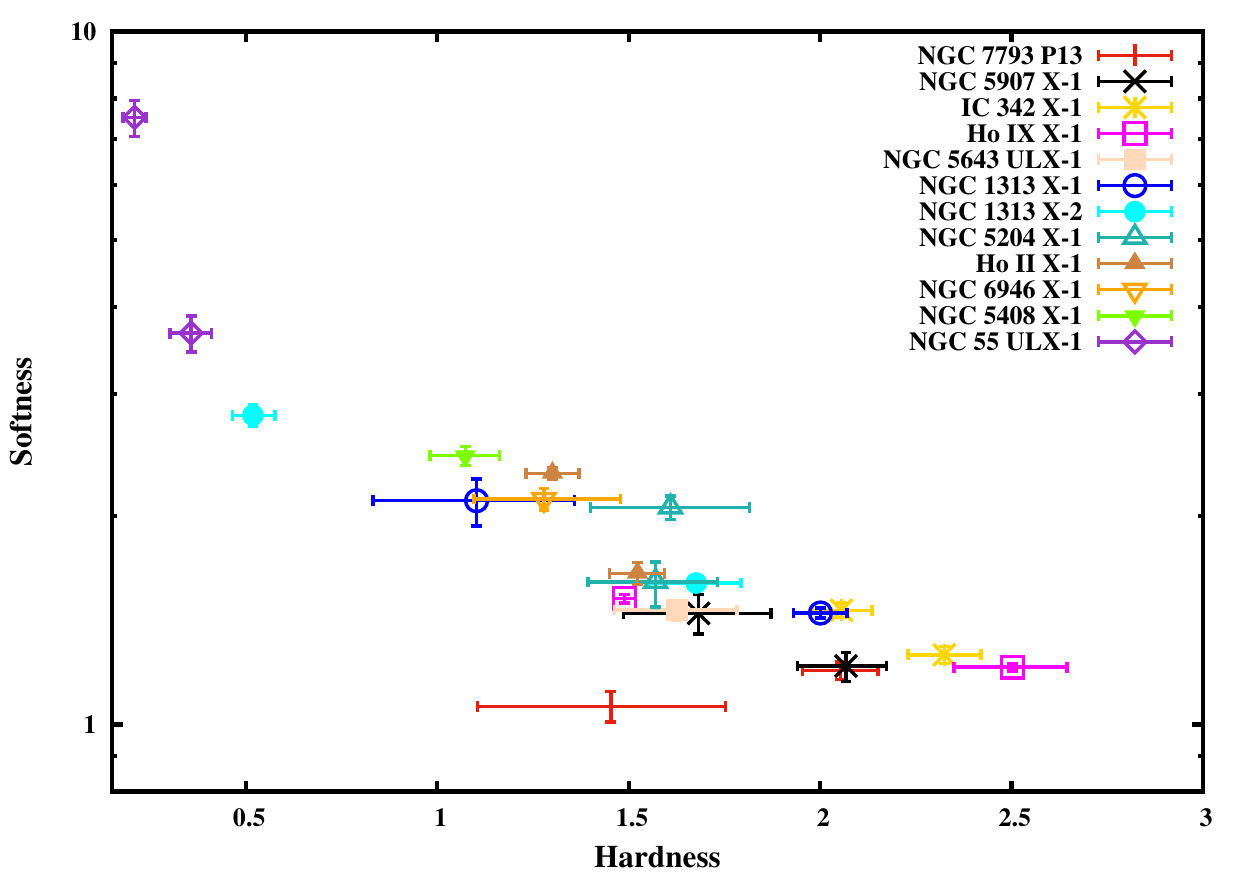}
\caption{Color-color diagram obtained from the ratios of the flux estimated in the energy ranges 2--4 keV, 4--6 keV and 6--30 keV and estimated from the best-fits of table~\ref{spectr}.}
\label{col-col}
\end{figure*}

We found that the ``pulsator-like'' model can describe well the spectra of virtually all the ULXs of our sample. Different broad band spectral shapes and best fit parameters characterize different sources (see Fig.~\ref{fig_spectra} and Table~\ref{spectr}). 
This is particularly evident at low energy, where, besides the effect of different absorption column densities  (from   $\sim5\times10^{20}$ cm$^{-2}$  to $8\times10^{22}$ cm$^{-2}$), the relative importance of the soft component can differ significantly (even if similar temperatures around  kT$\sim$0.2 keV are found). 
Spectral differences are apparent also at higher energies, where the spectrum is dominated by the cut-off power-law component. In order to study in greater detail differences in the harder spectral component (where the effects due to absorption and the additional soft component are negligible), search for correlations with other source properties, and possibly gather information on the innermost regions of the accretion flow close to the compact object, we characterize the high-energy spectra through a color-color diagram. This was done by using the fluxes in the  2--4 keV, 4--6 keV and 6--30 keV energy bands derived from the best-fit of Table~\ref{spectr}. We present the color-color diagram in figure~\ref{col-col}, where we defined the {\it hardness} as the (6--30 keV)/(4--6 keV) ratio and the {\it softness} as (2--4 keV)/(4--6 keV) ratio. { We note that the fluxes above 10 keV are estimated directly from the spectra only in the cases where {\it NuSTAR} observations are available. In the other cases, the fluxes are calculated from an extrapolation of the 0.3--10 keV best-fit model.}
As expected, sources displaying spectral variations move across different regions in this diagram 
(note again that we  pre-selected the most different spectra of each source) 
Several of the sources with parameters around {\it hardness}$\sim$1.5 and {\it softness}$\sim$1.5  become either softer (i.e. move toward the upper-left corner of the figure) or harder (i.e. move toward the lower-right corner),  but none of them shows both behaviors. 
This leads to a distinction between two different classes of softer and harder ULXs. A similar conclusion was   reached in e.g. \citet{sutton13} and \citet{pintore14} based on a different spectral analysis.

We highlight that the two known ULX pulsars in the sample lie in the region of the  color-color diagram that corresponds to the hardest spectra. Their {\it hardness} values are close to the maximum but still within the broad distribution of the ULXs in our sample.
{ We applied a Kolmogorov-Smirnov test (KS) to the distribution of the {\it softness} of 
pulsating and non-pulsating ULXs finding a KS-statistics of $\sim88\%$ and a {\it p}-value of $\sim0.013$ that they are drawn from the same sample. } 
Although with the current limited sample it is difficult to obtain firm conclusions, it is interesting to note that also the ULX pulsar M82 X-2 has a rather hard spectrum\footnote{Unfortunately, good quality \xmm\ spectral data are not available for the ULX pulsar M82 X-2, owing to contamination from the bright nearby ULX M82 X-1.}  \citep{brightman16b,brightman16a}.

We note that NGC 7793 P13 attained the lowest values of the {\it softness} ratio of the whole sample.
During the epoch 1 observation, when no pulsations were found in the timing data \citep[][a]{israel16a}, the spectrum of NGC 5907 X-1 was close to the center of the distribution in Fig.~\ref{col-col}. In 2014, when pulsations were detected, the  position of NGC 5907 X-1 in the color-color diagram was closer to that of NGC 7793 P13.
Hence, when pulsations were seen, the two ULX pulsars were significantly hard. The only sources that, at least on one epoch, had spectral parameters similar to those of the ULX pulsars are IC 342 X-1 and Ho IX X-1.

\section{Discussion}
\label{discussion}

In previous studies, ULX spectra characterized by a curvature above 3--5 keV and a soft excess below $\sim$1 keV were described in terms of physically-motivated spectral models for accreting BHs of different masses \citep[e.g.][]{stobbart06,gladstone09,pintore12,sutton13,bachetti13,walton13}. Inspired by the recent discovery of pulsations in three ULXs \citep[][]{bachetti14,israel16b,israel16a}, we applied here a  model that has been commonly used to fit the X-ray continuum of accreting magnetic NSs \citep[e.g.][]{white95} to good quality \xmm\ and \nus\ spectra of a sample of bright ULXs. Such model, {\sc highecut$\times$powerlaw} plus a soft blackbody, was introduced empirically \citep[see][ and references therein]{white95} and later found to be similar to the spectra calculated from models of bulk and thermal Comptonization in magnetic accretion models \citep{becker07}  which have also been employed to fit the spectra of a few X-ray pulsars in HMXBs \citep{farinelli16}.
The shape of its exponential cutoff is regulated by two parameters (the e-folding energy 
$E_f$ and the cutoff energy $E_c$), through which the spectra with vastly different curvature observed 
in accreting X-ray pulsars at high energies (usually $>7-10$~keV) can be fitted.   
In this work we found that this ``pulsator-like'' continuum model, together with a soft component, 
can be successfully used to describe also the overall shape of
the spectra of bright ULXs\footnote{As with other continuum models for ULXs, in a few cases 
the addition of broad absorption-like features \citep[e.g.][]{pinto16} and/or a high energy tail \citep[e.g.][]{walton14,mukherjee15} proved necessary in order to make the fit acceptable}.
This suggests also that accreting BH-based models usually adopted for ULXs (e.g. disc blackbody plus Comptonization) do not necessarily have the interpretative power that has been attributed to them.

In our spectral fits, the soft blackbody component has a temperature of $\sim$0.1-0.3 keV in all sources, but its relative importance differs across different spectra and sources. The equivalent radius of the thermally emitting region, inferred from the normalization of the blackbody component, is typically larger than 100 km. 
This value rules out any association with the putative NS surface, but may be consistent with the size of the emitting region of the disc close to the co-rotation radius, as in the case of some X-ray pulsators and ULX pulsars \citep[][]{israel16b,israel16a}. Alternatively, since pulsating ULXs have super-Eddington accretion rates, the soft component may originate in a powerful outflow that is ejected from the accretion disc \citep[e.g.][]{poutanen07,ohsuga11,middleton15}.

The best fit  values derived with the {\sc highecut $\times$powerlaw} for the hard component span a fairly large range, both in spectral slope and cut-off and folding energies, as observed also in galactic accreting X-ray pulsars. 
This component is believed to arise in the post-shock region of the accretion column above the surface of the neutron star. Recent analytical calculations, as well as 2D radiation-hydro-simulations of accretion onto  magnetized NSs, show that for magnetic fields of $\sim 10^{14}$~G or larger, very 
high luminosities  (up to 2-3 orders of magnitude above the Eddington limit) can be produced \citep{mushtukov15,kawashima16}. 

Based on currently available data, it is not possible to assess whether BH-models or
the pulsator-like model we adopted provide a statistically better fit to the X-ray continuum
of bright ULXs. The two types of models resemble each other in that they 
both comprise a soft component (a blackbody or a disk-blackbody, which is very difficult to
distinguish at the soft X-ray energies where it is seen) and a harder component comprising
a powerlaw-like spectrum with exponential cutoff, which likely originates from Comptonization in an accretion disk corona in the case of BHs and from the post-shock region of a magnetically-funneled accretion column in the case of NSs. The shape of the cut-off can be smoother or steeper than a simple exponential in the pulsator-like model. In order to ascertain what observations would be required to 
distinguish between BH and NS-like spectral models we simulated the spectra of IC 342 X-1 and NGC 1313 X-1 by using the best-fit {\sc diskbb+comptt} model (see e.g. \citealt{middleton15}) and a 1~Ms exposure time with {\it XMM-Newton} and {\it NuSTAR}. By fitting these spectra with a {\sc highecut$\times$powerlaw} model, we found structured spectral residuals at high energy ($>10$ keV), not observed with a {\sc diskbb+comptt} model.  The same simulation  was carried out based on the specifications of the future X-ray mission {\it Athena+}\footnote{http://www.the-athena-x-ray-observatory.eu/} where a discrimination of the two models required an exposure time of at least $\sim100$ ks in the WFI instrument. { However, we highlight that the simple {\sc diskbb+comptt} model, when using {\it NuSTAR} data, has been  challenged in the sources Ho IX X-1 \citep{walton14}, already indicating that ULX spectra do not completely suit with a BH description only. }

The relatively large number of parameters in the {\sc bb + highecut$\times$powerlaw} model hampers a straightforward comparison between the sources, but a clear picture emerges from a color-color diagram 
in which only the range above 2 keV is considered, 
where absorption effects are negligible and there is no (or marginal) contamination from the soft component. 
This clearly shows that the ULX pulsars, when pulsations are detected, have a  harder spectrum  than that of the majority of the other ULXs of the sample (note that the spectrum of NGC 5907 X-1 was somewhat softer in 2013,  when no pulsations from this source were detected). 
There are no instrumental effects that  favor the detection of pulsations, if present, in  harder sources, but  we cannot exclude that for some processes intrinsic to the sources the flux modulation   increases when their spectra become harder. { It is possible to speculate that the harder sources are those observed more directly in the inner regions close to the compact objects (or possibly from an accretion column, \citealt{fuerst16}) and this might facilitate the detection of pulsations in the case of accreting NS; while for softer sources the inner regions might be possibly occluded (because of higher angles of the line of sight) by cold, optically thick plasmas which reduce the pulsation coherence.}

None of the other sources showed high-energy spectra as hard as the hardest spectra of the ULX pulsars, except for IC 342 X-1 and Ho IX X-1. The spectral parameters of these two sources overlap, at least on one epoch, with those of the ULX pulsars (see Fig.~\ref{col-col}). Hence we suggest that  IC 342 X-1 and Ho IX X-1 may be good candidates NS ULXs. On the basis of their spectrum alone,  NGC 5408 X-1, Ho II X-1, NGC 1313 X-1 and X-2, NGC 5643 ULX1 and NGC 6946 X-1 are instead less likely to contain NSs.
The failure to detect pulsations so far in IC 342 X-1 and Ho IX X-1 may be caused by intermittent activity of  periodic flux modulation and/or by the presence of a large period derivative. However it should be also noted that X-ray luminosity swing of the ULX pulsars is much larger ($\ga$ a factor of 100) than that of IC 342 X-1 and Ho IX X-1 (a factor of $\sim 5$ only). Following Israel et al. (2016a,b) and assuming that the sources are all viewed at comparable inclination angles and have comparable beaming, we  expect that the higher the luminosity the higher the magnetic field. Israel et al. (2016a,b) estimated a dipolar magnetic field of $\sim3\times10^{13}$ G for NGC 5907 X-1 and $\sim5\times10^{12}$ G for NGC 7793 P13. Since IC 342 X-1 and Ho IX X-1 have luminosities of $\sim10^{40}$ erg s$^{-1}$, we can roughly guess a dipolar magnetic field of $\sim$$10^{13}$ G for both sources if their spin period will turn out to be in the range of seconds as well. 

\section{Conclusion}
We adopted a commonly used spectral model for accreting X-ray pulsating NS in the Galaxy to describe the spectra of a sample of bright ULXs. We found that this model well describes the spectra of most ULXs, suggesting that spectral models based on accreting BHs may not have the interpretative power that has been ascribed to them up to now. We conclude that other ULXs may be found to host NSs accreting at super-Eddington rates. We showed also that the known pulsating ULXs are amongst the hardest sources of our sample, with IC 342 X-1 and Ho IX X-1 having comparably hard spectra. Based on their spectral properties alone, we thus suggest these two sources are likely candidate NS ULXs.

\section*{Acknowledgements} 
We thank the anonymous referee for his/her helpful comments. We acknowledge financial contribution from the agreement ASI-INAF I/037/12/0 and PRIN INAF 2014.

\end{document}